\documentclass[aps,prl,twocolumn,superscriptaddress]{revtex4-1}

\usepackage{graphicx}
\usepackage{epstopdf}
\usepackage{color}
\usepackage{epsfig}

\usepackage[english]{babel}
\usepackage{amsmath}
\usepackage{amssymb}
\usepackage{natbib}

\newcommand{\LGF}{$\rm LiGdF_4$}
\newcommand{\LYGF}{$\rm LiY_{1-x}Gd_xF_4$}

\begin{document}
\author{S.\,S.\,Sosin}
\email{sosin@kapitza.ras.ru}
\affiliation{P. Kapitza Institute for physical problems RAS, 117334 Moscow, Russia}
\affiliation{HSE University, 101000 Moscow, Russia}
\author{A.\,F.\,Iafarova}
\affiliation{P. Kapitza Institute for physical problems RAS, 117334 Moscow, Russia}
\affiliation{HSE University, 101000 Moscow, Russia}
\author{I.\,V.\,Romanova}
\affiliation{Kazan Federal University, 420008 Kazan, Russia}
\author{O.\,A.\,Morozov}
\affiliation{Kazan Federal University, 420008 Kazan, Russia}
\author{S.\,L.\,Korableva}
\affiliation{Kazan Federal University, 420008 Kazan, Russia}
\author{R.\,G.\,Batulin}
\affiliation{Kazan Federal University, 420008 Kazan, Russia}
\author{M.\,Zhitomirsky}
\affiliation{Institut Laue-Langevin, 38042 Grenoble, France}
\author{V.\,N.\,Glazkov}
\email{glazkov@kapitza.ras.ru}
\affiliation{P. Kapitza Institute for physical problems RAS, 117334 Moscow, Russia}
\affiliation{HSE University, 101000 Moscow, Russia}

\title{Microscopic spin Hamiltonian for a dipolar-Heisenberg magnet \LGF\ from EPR measurements}
\date{\today}

\begin{abstract}
Low-temperature electron paramagnetic resonance measurements are performed on single crystals of LiY$_{1-x}$Gd$_x$F$_4$ with weak $x=0.005$ and moderate $x=0.05$ concentration of Gd ions. Modeling of the experimental spectra allows us to precisely determine microscopic parameters of the spin Hamiltonian of the parent \LGF\ material, including the nearest-neighbor exchange constant. The obtained parameters are further tested by comparing a strongly anisotropic Curie-Weiss temperature obtained for \LGF\ in our static magnetization measurements with theoretically computed values. We find a fine balance between principal magnetic interactions in \LGF, which results in a hidden magnetic frustration presumably leading to a delayed magnetic ordering and an enhanced magnetocaloric effect at low temperatures.
\end{abstract}
\maketitle

{\it Introduction.}\ \ Lithium rare-earth fluorides LiREF$_4$ is a family of magnetic materials with dominant dipolar interactions. However, their magnetic behavior, including the type of magnetic ordering, is significantly influenced by single-ion properties of magnetic rear-earth ions. For the case of a strong easy-axis (Ising-type) anisotropy of Ho$^{3+}$ ions in LiHoF$_4$ the dipolar interaction stabilizes ferromagnetic structure with
the temperature of magnetic ordering $T_C =1.53$~K approximately corresponding to the dipolar field strength~\cite{cooke}. Critical properties of this material in an applied field ${\bf H}\perp c$ were extensively studied from the point of view of a realization of a paradigmatic transverse-field Ising model \cite{bitko,Ronnow1,Ronnow2,Kovacevic}. For erbium compound, magnetic moments of Er$^{3+}$ ions exhibit a strong planar anisotropy and
the magnetic ordering is antiferromagnetic with significantly lower transition temperature $T_N = 0.38$~K \cite{beauvillain3}. This system was recently studied in more detail including the magnetic structure and the critical behavior under applied field~\cite{kraemer}. The enhanced role of fluctuations due to the intrinsic frustration of the long-range dipolar interaction in this geometry was suggested.

The most isotropic material in this family is LiGdF$_4$ with Gd$^{3+}$ ions in the $s$-state with the spin-only angular momentum $S=7/2$. This material was recently recognized as an excellent refrigerant material for the low-temperature magnetic cooling \cite{Numazawa06}. Despite its popularity for the refrigeration applications \cite{Shirron04,Numazawa09,Wikus14}, there is an apparent lack of knowledge on its basic magnetic properties. In particular, no magnetic ordering was observed so far down to temperatures 0.3-0.4~K~\cite{Numazawa06,Wikus14,Babkevich15}. The delayed magnetic ordering can presumably originate from a fine balance of dipolar and exchange interactions taking into account a relatively weak (compared to the other two above mentioned systems) single-ion anisotropy. Analogously to magnets with frustrated exchange interaction (see {\it e.g.} \cite{gardner_review} and references therein), this balance may be a prerequisite for the presence of a number of exotic magnetic phases and peculiar phase transitions under applied field. In combination with a relatively high density of magnetic ions with nearly degenerate eightfold eigenstate (about $1.34\cdot 10^{22}/{\rm cm}^3$), comparable to those in other Gd-based magnetically frustrated compounds ({\it e.g.} $1.26\cdot 10^{22}/{\rm cm}^3$ for $\rm Gd_3Ga_5O_{12}$ and $1.52\cdot 10^{22}/{\rm cm}^3$ for $\rm Gd_2Ti_2O_7$) the delayed magnetic ordering also provides a unique opportunity for practical applications in the field of magnetic refrigeration at ultra-low temperatures~\cite{sosin}.

This work is aimed at precise determination of the spin-Hamiltonian parameters of \LGF\ by experimental study and modeling of electron paramagnetic resonance (EPR) spectra of the isostructural nonmagnetic yttrium compound with small and moderate concentration of Gd-ions (\LYGF\ with $x=0.005$ and 0.05). The obtained values of parameters are shown to describe well the anisotropy of a Curie-Weiss temperature (CWT) observed in our static magnetization measurements for a concentrated \LGF\ compound.

\begin{figure}
\centering
\includegraphics[width=0.8\columnwidth]{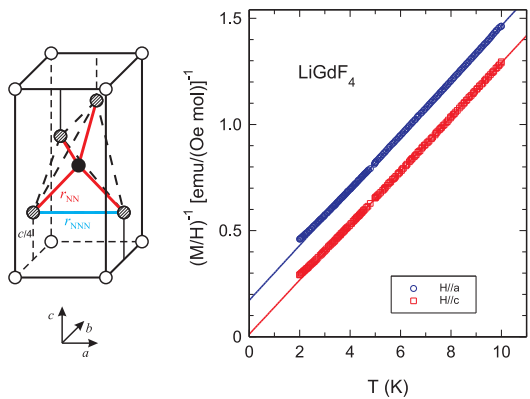}
\caption{Fig.~\ref{structure-magn}. (Left panel): The tetragonal unit cell of LiREF$_4$ compounds (only RE-sites are shown), hatched circles correspond to four nearest neighbors forming the tetrahedron (dashed lines) around the central (black) site. (Right panel): Temperature dependence of the inverse molar susceptibility of \LGF\ measured in a field applied along $a$- and $c$-axis (symbols {\footnotesize $\bigcirc$} and $\Box$ respectively). Linear fits to high-temperature (above 6~K) parts of the data with the parameters quoted in the text, are shown by solid lines.}
\label{structure-magn}
\end{figure}

{\it Crystal structure and growth}: The compounds of LiREF$_4$ family (RE is a trivalent rare-earth ion) have Scheelite-type structure with the space group $I4_1/a$ (C$_{\rm 4h}^6$) and the local symmetry S$_4$ on each RE-site. The tetragonal unit cell shown in Fig.~\ref{structure-magn} (left panel) contains four formula units. It has the following parameters: $a=5.219$ and $c=10.97$~{\AA} for the Gd compound and $a=5.175$ and $c=10.74$~{\AA} for the isostructural yttrium-based compound~\cite{Keller}. The corresponding Bravais lattice is a body-centered tetragonal lattice with a primitive unit cell consisting of two RE-ions. Each ion in a RE-site is surrounded by four nearest neighbors (NN) located at a distance $r_{\rm NN}=\sqrt{(a/2)^2+(c/4)^2}\simeq 3.79$~{\AA} and four next-nearest neighbors (NNN) in an $(ab)$-plane at a distance $r_{\rm NNN}=a$ (see Fig.~\ref{structure-magn}, left panel).

All single-crystal samples studied in this work were grown using a standard Bridgman-Stockbarger technique. The directions of crystal axes were precisely determined by X-ray Laue diffraction patterns.

{\it Static magnetic susceptibility}: Magnetization of \LGF\ was measured using the Quantum Design PPMS Vibrating Sample Magnetometer. The sample was cut from the parent single crystal in a shape of a thin plate 16.8~mg by mass containing the $ac$ crystal plane. A weak magnetic field $H$ has been applied along the two principal crystal axes, $c$ and $a$, within the sample plane to exclude the demagnetization corrections. The temperature of the experiment varied from 2 to 10~K with the data obtained on cooling and heating being indistinguishable. The experimental data are presented in Fig.~\ref{structure-magn} (right panel) showing the inverse magnetization to field ratio vs temperature. Linear fits to the data above 6~K according to the paramagnetic Curie-Weiss law $M=(g\mu_B)^2S(S+1)H/[3k_B(T-\theta_{CW})]$
($\mu_B$ is a Bohr magneton, $k_B$ is a Boltzmann constant, $\theta_{CW}$ is a Curie-Weiss temperature) shown by solid lines allowed us to determine both the corresponding $g$-factors and CWTs for both field directions: $g^a=1.99 (1)$, $g^c=2.00 (1)$, $\theta_{CW}^a=-1.33 (5)$~K and $\theta_{CW}^c=-0.08 (5)$~K. An unusually large relative anisotropy of the CWT provides another evidence for a quite atypical compensation of all substantial interactions in the system, once again challenging for precise determination of microscopic spin-Hamiltonian parameters.

\begin{figure}
\centering
\includegraphics[width=0.9\columnwidth]{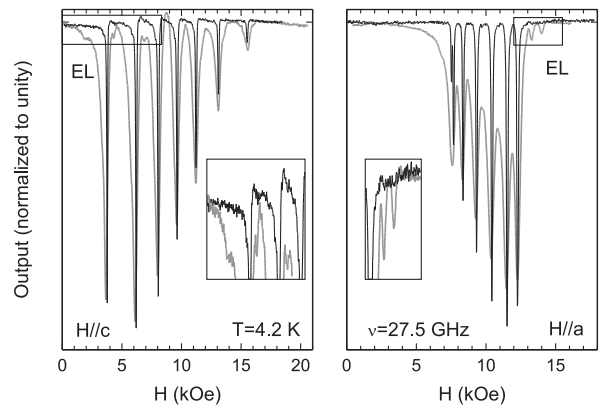}
\caption{Fig.~\ref{spectra_compare}. Resonance absorption spectra in \LYGF\ samples ($x=0.005$ -- thin lines, $x=0.05$ -- bold lines) recorded at a frequency $\nu=27.5$~GHz and temperature $T=4.2$~K: $H\parallel c$ (left panel), $H\parallel a$ (right panel). Insets in both panels show expanded boxed areas with ``extra lines'' in the sample with $x=0.05$.}
\label{spectra_compare}
\end{figure}

\begin{figure}
\centering
\includegraphics[width=0.9\columnwidth]{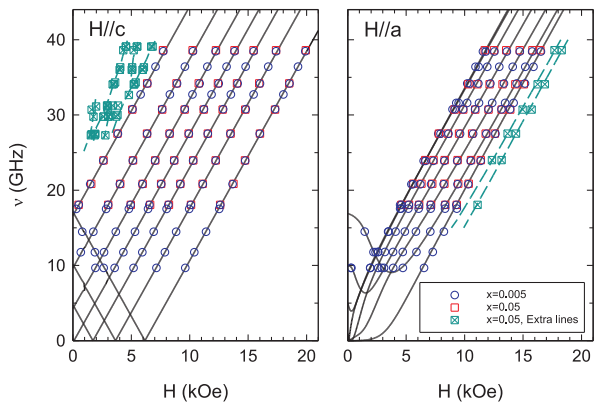}
\caption{Fig.~\ref{ffd}. Frequency-field diagrams of EPR-spectra in \LYGF\ with $x=0.005$ (main spectral components are shown by {\footnotesize $\bigcirc$}) and 0.05 ($\Box$), obtained at $T=4.2$~K for two principal orientations of the sample under external field. Symbols $\boxtimes$ in both panels correspond to ``extra lines'' existing in $x=0.05$ sample only. The positions of resonance absorption peaks calculated using the single-ion Hamiltonian~(\protect\ref{anisotropy}) with the parameters quoted in the text, are shown by solid lines; dashed lines correspond to the spectra of the NN pairs coupled by exchange ($J_{\rm NN}$) and dipolar interactions.}
\label{ffd}
\end{figure}

\begin{figure}
\centering
\includegraphics[width=0.9\columnwidth]{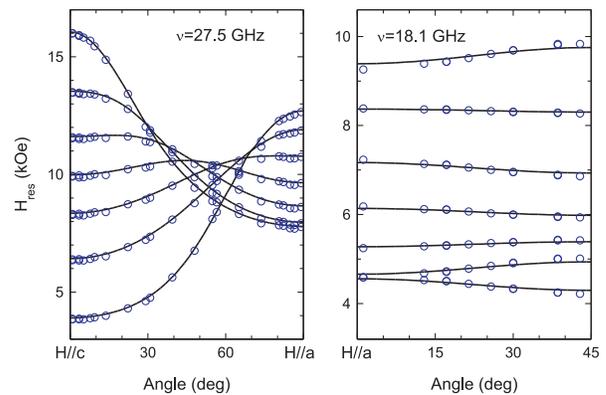}
\caption{Fig.~\ref{angular}. Angular dependence of seven main spectral components of the $S=7/2$ multiplet in \LYGF\ ($x=0.005$) sample rotated from $H\parallel c$ to $H\parallel a$ (left panel) and within the tetragonal plane (right panel). The resonance fields are shown by circles, solid lines are the result of the single-ion modeling.}
\label{angular}
\end{figure}

{\it Single-ion EPR}: In the following we describe the procedure of the exact determination of spin-Hamiltonian parameters in \LGF\ based on experimental study and modeling of EPR-spectra in \LYGF\ single crystals, a system with an isostructural non-magnetic matrix weakly doped by magnetic Gd-ions ($x=0.005$). The samples were in a shape of thin rectangular plates cut in an $ac$- or tetragonal plane with the size varying from $1\times 1\times 0.2$ to $2\times 2\times 0.2$~mm$^3$ (approximately 1 to 5~mg by mass). The experiments were performed using a set of home-made transmission-type microwave spectrometers equipped with multi-mode rectangular cavities (resonators) with eigen-frequencies starting from 9~GHz and higher. Resonators were placed inside an inner vacuum space of a $^4$He-bath cryostat with the base temperature of 1.8~K. The variable-temperature stage was equipped with the heater and thermometer to stabilize temperature from 1.8 to 10~K with an accuracy better than $5\%$. In some experiments the sample was rotated using a worm-gear setup with an accuracy exceeding $1^{\circ}$. The microwave signal transmitted at a fixed frequency and temperature was detected and recorded on continuous back and force sweep of magnetic field up to 7~T created by a superconducting cryomagnet. The examples of these records in a \LYGF\ single crystal ($x=0.005$) obtained at a frequency $\nu =27.5$~GHz and temperature $T=4.2$~K are shown in Fig.~\ref{spectra_compare} for two principal directions of an applied field, $H\parallel c$ and $H\parallel a$. The spectra consist of seven main lines corresponding to transitions between energy sub-levels of $S=7/2$ eightfold multiplet splitted by single-ion anisotropy and external field. The results obtained at several frequencies are summarized in corresponding frequency-field diagrams (the positions of lines are shown in Fig.~\ref{ffd} by circles). Rotation of the sample from $H\parallel c$ to $H\parallel a$ directions results in continuous shifting of spectral lines demonstrated on the angular dependence (Fig.~\ref{angular}, left panel). We have also observed a small 90-degree periodic shift of the resonance peaks when rotating the sample with an external field applied in tetragonal plane indicating the presence of a weak forth-order in-plane anisotropy (Fig.~\ref{angular}, right panel).

The EPR-spectra have been simulated via exact diagonalization of the $S=7/2$ single-ion spin-Hamiltonian in which we keep three lowest-order invariants:
\begin{equation}
\hat{\cal H}_{\rm SI}= D_2 S_z^2 + D_4S_z^4+\frac{E}{2}(S_x^2 S_y^2 +S_y^2 S_x^2) +
\mu_B g^{\alpha\beta} H^\alpha S^\beta \,.
\label{anisotropy}
\end{equation}
Positions of measured resonance lines for all field directions (solid lines in Figs.~\ref{ffd},\ref{angular}) are well reproduced for the isotropic $g$-tensor $g^a=g^c=1.984(14)$ and the following set of the anisotropy constants:
$$
D_2/k_B=-0.096 (2)~{\rm K}\,,~~D_4/k_B =-0.0018 (6)~{\rm K}\,,
$$
$$
E/k_B =0.0020 (2)~{\rm K} \,.
$$
Signs of the crystal-field parameters are determined from relative intensities of spectral lines, which were calculated taking into account matrix elements of transitions between corresponding sub-levels and their thermal population. Parameters of the six-order anisotropy terms that are generally allowed for $S=7/2$ spins in the $S_4$ local symmetry cannot be reliably determined from our data. For the single-ion Hamiltonian expressed in terms of the conventional Stevens operators, see {\it e.g.}\ \cite{AABB}, these values can be directly matched to corresponding coefficients:
$$
b_{20}=-2.619 (60)~{\rm GHz}\,,~~b_{40}=-0.058 (18)~{\rm GHz}\,,
$$
$$
b_{44}=0.306 (30)~{\rm GHz} \,.
$$

The obtained values are in reasonable agreement with the results reported in~\cite{Misra,Misiak,aminov}.

{\it Spectra of exchange and dipolar coupled pairs}: The exchange constants were determined by studying EPR-spectra in \LYGF\ samples with moderate concentration of magnetic ions ($x=0.05$). A set of seven main lines was also detected with their positions and relative intensities being nearly identical to those observed in a sample with smaller Gd concentration (the positions are shown in Fig.~\ref{ffd} by squares). However, the lines observed in $x=0.05$ sample demonstrated considerable dipolar broadening: the half width at half maximum of the lines in $x=0.005$ sample varies in the range $\Delta H_{\rm HWHM}=30\div 50$~Oe, while that in $x=0.05$ sample ranges from 200 to 300~Oe. In addition to main spectral components, the spectra in this sample contain much less intensive resonance lines which are either fully absent or negligibly weak in the sample with $x=0.005$. These lines marked in Fig.~\ref{spectra_compare} as ``EL'' (extra lines), can be presumably identified as the spectrum of pairs of magnetic ions coupled by exchange and dipolar interactions since the concentration of these pairs is quadratic in $x$. The positions of the most clearly visible and reliably determined lines are shown in the general frequency-field diagram (Fig.~\ref{ffd}) by symbols $\boxtimes$.

We have performed a series of EPR-spectra simulations analogous to those described above, for two $S=7/2$ Gd$^{3+}$ ions with the spin-Hamiltonian including Heisenberg exchange and dipolar coupling, as well as the single-ion contribution and magnetic field.
\begin{eqnarray}
\hat{\cal H}_{12} & = & \sum_{i=1}^2\hat{\cal H}_{\rm SI}({\bf S}_i) + J\,\mathbf{S}_1\cdot \mathbf{S}_2  \nonumber \\
& + & (g\mu_B)^2\left[\frac{{\bf S}_{1}\cdot {\bf S}_{2}}{{\bf r}_{12}^3}-
\frac{3({\mathbf S}_{1}\cdot {\bf r}_{12})({\bf S}_{2}\cdot{\bf r}_{12})}{{\bf r}_{12}^5}\right].
\label{12}
\end{eqnarray}
\begin{figure}
\centering
\includegraphics[width=0.7\columnwidth]{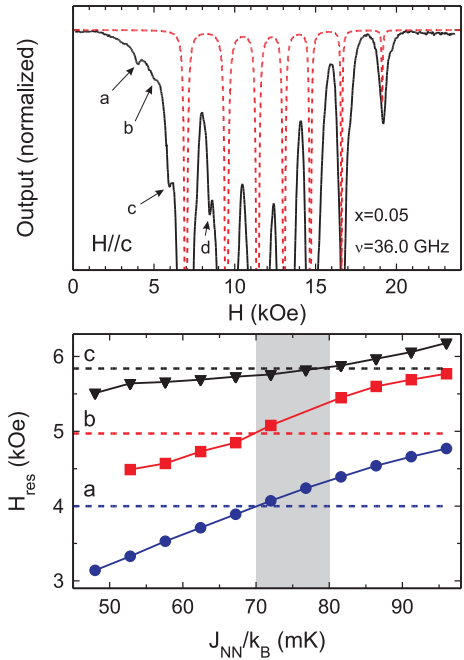}
\caption{Fig.~\ref{nn}. (Upper panel): Fragment of the resonance absorption curve recorded in \LYGF\ ($x=0.05$) sample at a frequency $\nu =36.0$~GHz for $H\parallel c$ direction of the applied field; the best visible ``Extra lines'' are marked by arrows, dashed line is an EPR spectrum simulated within the single-ion model. (Lower panel): simulated positions of a,b,c-lines depending on the value of the nearest-neighbor exchange interaction, horizontal dashed lines mark their real positions, shadowed area is an interval of allowed $J_{\rm NN}$ values.}
\label{nn}
\end{figure}
\noindent
Both the nearest- and the next-nearest-neighbor spin pairs with exchange constants $J_{\rm NN}$ and $J_{\rm NNN}$, respectively, and corresponding dipolar energies were considered since both types of pairs are equally probable for \LGF\ structure. (Note that some of the pairs become nonequivalent as the magnetic field deviates from the tetragonal axis). Single-ion anisotropy parameters and $g$-factors are determined from single-ion spectra as was described above, the dipolar coupling is predetermined by the crystal structure. Thus, the model includes only two independent tunable parameters, $J_{\rm NN}$ and $J_{\rm NNN}$.

Simulated EPR spectra of two coupled $S=7/2$ ions contain numerous absorption lines with different intensities. However, most of them are located in the field range dominated by strong single-ion absorption components. Hence, we focus our attention at spectral components which are well separated from the single-ion spectra. Some of them, in particular, low-field absorption lines observed at $H\parallel c$, were found to be highly sensitive to the values of coupling parameters. The examples of these lines are shown in the upper panel of Fig.~\ref{nn} (marked as a,b and c). We found that the positions of these components can be reproduced solely in the NN model. Varying the exchange coupling parameter $J_{\rm NN}$ and tracing the corresponding shifts of a,b and c resonance fields (Fig.~\ref{nn}, lower panel), one can determine the allowed interval for $J_{\rm NN}$ values shown in Fig.~\ref{nn} by shadowed area: 0.07~K$\lesssim J_{\rm NN}/k_B\lesssim 0.08$~K or $J_{\rm NN}=1.55 (10)$~GHz. Positive sign of $J$ implies this interaction to be antiferromagnetic. The rest of the observed spectral components which are missed in NN modeling ({\it e.g} line ``d'' in Fig.~\ref{nn}) can be identified as the spectrum of NNN pairs. However, the value of the NNN exchange interaction could not be reliably determined from our modeling since the positions of all detected NNN components appeared to be practically independent on $J_{\rm NNN}$ while the others are hidden under intense single-ion spectral lines.

\begin{figure}
\centering
\includegraphics[width=\columnwidth]{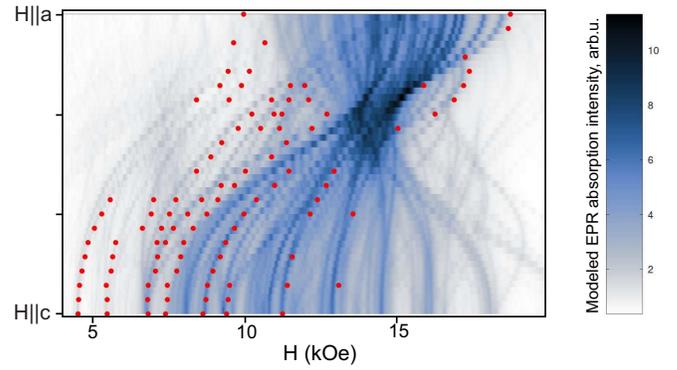}
\caption{Fig.~\ref{angular_pair}. Experimental angular dependence of the resonance fields for the spectral components identified as originating from the coupled pairs (symbols) and simulated paramagnetic absorption pattern for $\nu =39.0$~GHz, $c$ to $a$ rotation (grades of color). Note that the intermediate field area ($7\div 18$~kOe at this frequency) is of little use for detecting weak pair spectral lines since it is covered by much more intensive single-ion resonance modes.}
\label{angular_pair}
\end{figure}

To ultimately ensure that weak spectral components are properly identified as the absorption from coupled Gd ions we have measured their angular dependence at a fixed frequency and compared it with the results of simulation taking into account NN and NNN pairs for all possible orientations of pairs with respect to the applied magnetic field. In this procedure the parameters $J_{\rm NN}$ and $J_{\rm NNN}$ were set to 0.075~K and zero, respectively. Fig.~\ref{angular_pair} demonstrates satisfactory agreement between experiment and simulation for all components that can be reliably distinguished at the shadowing background of much more intensive single-ion spectral components: most of identified coupled-pair lines (closed symbols) follow predicted angular dependences. The frequency-field dependence of some of the most clearly visible lines was also calculated for two principal field directions $H\parallel c,a$ with the parameter $J_{\rm NN}/k_B=0.075$~K. The corresponding curves shown in Fig.~\ref{ffd} by dashed lines also reproduce quite well the experimental data.

{\it Theoretical analysis}: We now use our data for the temperature-dependent susceptibility in the bulk \LGF\ to validate the microscopic parameters obtained from the EPR study of dilute samples. In particular, we explain strongly anisotropic values of the Curie-Weiss temperature $\theta_{CW}$,
obtained from the high-$T$ limit of the magnetic susceptibility $\chi \propto 1/(T-\theta_{CW})$, as illustrated in Fig.~\ref{structure-magn}.

Different terms (exchange, single-ion and dipolar) in the spin Hamiltonian contribute additively into the CWT \cite{Jensen,Lhotel21}:
$\theta_{CW}^{\alpha}=\theta_{ex}+\theta_{\rm SI}^{\alpha}+\theta_{dd}^{\alpha}$, where $\alpha$ denotes the direction of an applied field. Corresponding contributions can be computed as $\theta_{CW}^{\alpha} = - C_2^{\alpha}/C_1$ with $C_1 = S(S+1)/3$ and
\begin{equation}
C_2^{\alpha} = -\frac{1}{N}\sum_{i,j}\langle (S_{i}^{\alpha}S_{j}^{\alpha}\rangle){\cal H}\rangle_c \,.
\label{C2}
\end{equation}
Here $N$ is the total number of magnetic ions and $\langle ...\rangle_c$ stands for a cumulant $\langle AB\rangle_c =\langle AB\rangle_0 -\langle A\rangle_0\langle B\rangle_0$, with $\langle ...\rangle_0$ being the paramagnetic average.

The isotropic exchange contribution is given by the standard expression
\begin{equation}
k_B\theta_{ex} = -\frac{1}{3}\left(z_{\rm NN}J_{\rm NN}+z_{\rm NNN}J_{\rm NNN}\right)S(S+1) \,,
\end{equation}
where $z$'s are respective coordination numbers. For \LGF\ with $S=7/2$ and $z_{\rm NN}=z_{\rm NNN}=4$ we obtain $k_B \theta_{ex}=-21(J_{\rm NN}+J_{\rm NNN})$. The single-ion anisotropy~(\ref{anisotropy}) yields contributions with opposite signs into the CWTs:
\begin{eqnarray}
&& k_B \theta_{\rm SI}^c  =  -2k_B \theta_{\rm SI}^a=\frac{(2S-1)(2S+3)}{15} \times \\
&& \mbox{} \times \left[-D_2-\frac{D_4}{7}(6S^2+6S-5)+\frac{E}{7}(S^2+S+5) \right]. \nonumber
\end{eqnarray}

Finally, the effect of dipole interactions is accounted for by
\begin{equation}
k_B\theta_{dd}^\alpha = -\frac{g\mu_B S(S+1)}{3N}\sum_{i,j}\frac{1}{|{\mathbf r}_{ij}|^3}
\left[1-\frac{3(r^{\alpha}_{ij})^2}{|{\bf r}_{ij}|^2} \right].
\label{dip}
\end{equation}
We use the Ewald's summation technique to compute the conditionally converging lattice sum in~(\ref{dip}), see~\cite{Lhotel21} for further details.
Exclusion of a singular term in corresponding calculations assumes implicitly a vanishing demagnetization factor $N_\alpha=0$. This is precisely the condition satisfied by the geometry of our experiment for both field directions along the $a$- and $c$-axes. As a result, we find for \LGF\
$$
\theta_{dd}^c = 1.019 {\rm K}\,,\quad \theta_{dd}^a = 0.580~{\rm K} \,.
$$
Combining all relevant contributions and using experimentally determined anisotropy constants and $\theta_{CW}^{c,a}$, one can calculate the exchange interaction parameters. From $\theta_{CW}^c=-0.08 (5)$~K we get $(J_{\rm NN}+J_{\rm NNN})/k_B=0.077 (3)$~K, whereas $\theta_{CW}^a=-1.33 (5)$~K yields $(J_{\rm NN}+J_{\rm NNN})/k_B=0.079 (3)$~K. These values fall confidently within the range deduced from the EPR measurements for the nearest-neighbor exchange constant $0.07\leq J_{\rm NN}/k_B\leq 0.08$~K. Consequently, the next-nearest-neighbor exchange parameter should be very small: $J_{\rm NNN}/k_B\lesssim 0.005$~K. One should mention, that the exchange interaction between Gd ions in \LGF\ appears to be substantially different from those evaluated for Ho ions in an Ising-type system LiHoF$_4$ (about 1~mK~\cite{Ronnow1}) and estimated for Gd-Yb pairs from EPR experiments on Gd-doped LiY$_{1-x}$Yb$_x$F$_4$ system as $0.18\pm 0.12$~K~\cite{Misiak}.

{\it Conclusions}: To summarize, a comprehensive electron paramagnetic resonance study of \LYGF\ single crystal samples with different concentrations of Gd ions ($x=0.005$ and 0.05) supplemented by static magnetization measurements in the parent magnetic compound \LGF\ allow us to determine, precisely and self-consistently, all spin-Hamiltonian parameters in this system including $g$-factors and single-ion anisotropy constants as well as nearest- and next-nearest exchange interactions. The strength of all substantial magnetic interactions in the system appear to be of the same order of magnitude. The characteristic dipolar energy can be estimated as $E^{dd}= (g\mu_BS)^2/r_{\rm NN}^3\simeq 0.6$~K, the nearest-neighbor exchange interaction is equal to $E^{ex}_{\rm NN}=J_{\rm NN}S^2\simeq 0.9$~K, while the effect of a single-ion anisotropy is basically reduced to splitting between two lowest ionic sub-levels, so that $|S_z\rangle =\pm 7/2$ and $\pm 5/2$. Its value is directly measured in our experiment corresponding to an upper gap in the frequency-field diagram (Fig.~\ref{ffd}) which is approximately equal to 17~GHz ($\simeq 0.8$~K). Thus, the obtained results provide strong evidence for the fine balance between various types of magnetic interactions in \LGF\ that might lead to the delayed magnetic ordering and enhanced magnetocaloric effect, as well as supposedly generate peculiar phase diagram in this system in the low-temperature range.

\noindent
{\it Acknowledgments}: The authors thank V.\,A.\,Shustov for providing the results of X-ray measurements. The work (sample growth and EPR experiments) was supported by Russian Science Foundation, Grant No 22-12-00259. Data analysis (simulation of EPR spectra) was supported by Basic research program of HSE University. Magnetometry studies were supported by Kazan Federal University Strategic Academic Leadership Program (PRIORITY-2030).

\end{document}